\documentclass[a4paper,11pt]{article}
\pdfoutput=1 

\usepackage{jheppub} 

\usepackage[T1]{fontenc} 

\newcommand{\be}{\begin{equation}}
\newcommand{\ee}{\end{equation}}
\newcommand{\bea}{\begin{eqnarray}}
\newcommand{\eea}{\end{eqnarray}}
\newcommand{\nn}{\nonumber}
\newcommand{\eq}[1]{Eq.~\eqref{#1}}

\title{Power Corrections in the $N$-jettiness Subtraction Scheme}

\author[a]{Radja Boughezal,}
\author[b,c,d,e]{Xiaohui Liu,}
\author[f,a]{and Frank Petriello}

\affiliation[a]{High Energy Physics Division, Argonne National Laboratory, Argonne, IL 60439, USA}
\affiliation[b]{Department of Physics, Beijing Normal University, Beijing, 100875, China}
\affiliation[c]{Center of Advanced Quantum Studies, Beijing Normal University,Beijing, 100875, China}
\affiliation[d]{Center for High-Energy Physics, Peking University, Beijing, 100871, China}
\affiliation[e]{Maryland Center for Fundamental Physics, University of Maryland, College Park, Maryland 20742, USA}
\affiliation[f]{Department of Physics \& Astronomy, Northwestern University, Evanston, IL 60208, USA}
\emailAdd{rboughezal@anl.gov}
\emailAdd{xiaohuiliu@gmail.com}
\emailAdd{f-petriello@northwestern.edu}

\abstract{We discuss the leading-logarithmic power corrections in the $N$-jettiness subtraction scheme for higher-order perturbative QCD calculations.  We compute the next-to-leading order power corrections for an arbitrary $N$-jet process, and we explicitly calculate the power correction through next-to-next-to-leading order for color-singlet production for both $q\bar{q}$ and $gg$ initiated processes.  Our results are compact and simple to implement numerically.  Including the leading power correction in the $N$-jettiness subtraction scheme substantially improves its numerical efficiency.  We discuss what features of our techniques extend to processes containing final-state jets.
}

\begin{document} 
\maketitle
\flushbottom

\section{Introduction}
\label{sec:intro}

Accurate predictions for high-energy scattering processes at hadron colliders rely upon calculations to higher orders in the perturbative expansion of QCD.  Fully differential predictions are needed, in order to correctly model the final-state cuts imposed in all experimental analyses and directly compare theory with data.  The calculation of higher-order corrections is complicated by the fact that the real-emission and virtual corrections which contribute to the cross section exhibit infrared singularities that cancel only after they are combined.  At next-to-leading order (NLO) in the strong coupling constant several well-established techniques have been successfully applied for many years to accomplish this task~\cite{Catani:1996vz,Frixione:1995ms,Kosower:1997zr}.  In the past several years, new schemes~\cite{GehrmannDeRidder:2005cm, Somogyi:2005xz, Catani:2007vq, Czakon:2010td, Boughezal:2011jf,Cacciari:2015jma,Boughezal:2015dva,Gaunt:2015pea} have been proposed that enable calculations through the next-to-next-to-leading order (NNLO) in perturbative QCD, and permit precision comparisons of theoretical predictions with data from the Large Hadron Collider (LHC)

In this work we focus on the $N$-jettiness subtraction scheme for higher-order calculations~\cite{Boughezal:2015dva, Gaunt:2015pea}.  This conceptually appealing idea uses the $N$-jettiness event shape variable $\tau_N$~\cite{Stewart:2010tn} as a resolution parameter to isolate and cancel the double-unresolved singular limits, where two partons become soft and/or collinear, that complicate the calculation of NNLO cross sections.  The idea of using a physical observable to isolate singularities and construct subtraction terms at higher orders stems from the idea of $q_T$-subtraction~\cite{Catani:2007vq}, which uses the transverse momentum of a color-singlet object to accomplish NNLO computations for color-singlet processes.  This was further generalized to calculate top-quark decays~\cite{Gao:2012ja} and $t{\bar t}$ production in leptonic collisions~\cite{Gao:2014nva}.  The $N$-jettiness subtraction scheme further extends this idea to handle processes with arbitrary final-state jets.  When $\tau_N$ is large, an $N+1$ jet configuration is guaranteed, leading to a NLO contribution to the $N$-jet process that can be obtained with standard techniques.  When $\tau_N$ is small, the  NNLO results can be obtained using an effective theory approach~\cite{Bauer:2000ew,Bauer:2000yr,Bauer:2001ct,Bauer:2001yt,Bauer:2002nz}. The $N$-jettiness subtraction scheme has proven quite successful in enabling NNLO phenomenology.  It has led to some of the first calculations for vector boson production in association with a jet~\cite{Boughezal:2015dva, Boughezal:2015ded, Boughezal:2016yfp,Boughezal:2016dtm,Boughezal:2016isb} and Higgs production in association with a jet~\cite{Boughezal:2015aha} at the LHC through NNLO.  It has also led to first predictions for inclusive jet production at NNLO in electron-nucleon collisions~\cite{Abelof:2016pby}, and has reproduced known results for color-singlet production through NNLO~\cite{Gaunt:2015pea, Campbell:2016jau, Campbell:2016yrh, Boughezal:2016wmq}  Color-singlet production through NNLO using the $N$-jettiness subtraction scheme has been publicly released in the numerical code {\tt MCFM v8.0}~\cite{Boughezal:2016wmq}.

The $N$-jettiness subtraction scheme relies upon the introduction of a cutoff $\tau_N^{cut}$ that separates the $N+1$ jet configuration from the doubly-unresolved limit.  The below-cut region is expanded in $\tau_N^{cut}/Q$, where $Q$ denotes the hard momentum transfer in the process, in order to allow for an effective field theory calculation.  The cutoff must be chosen small so that the power corrections in $\tau_N^{cut}/Q$ are negligible.  However, the below-cut and above-cut contributions separately depend on logarithms of $\tau_N^{cut}/Q$ that only cancel after the two regions are combined.  Since these regions live in different phase spaces and are numerically integrated separately, these logarithms introduce numerical noise that challenge the efficiency of the method.  Although the numerics can already be controlled sufficiently for phenomenological applications, it is desirable for computational efficiency to reduce the sensitivity of the method to the power corrections.  An explicit calculation of at least the leading power correction would allow $N$-jettiness subtraction to be used with larger $\tau_N^{cut}$, reducing the computational cost of the approach.
 
 In this work we discuss the analytic calculation of the dominant power corrections through NNLO\footnote{Initial numerical results for these corrections have already been presented in Refs.~\cite{Boughezal:2016wmq} and~\cite{loopfesttalk}.}.  In Section~\ref{sect:nlo} we show in detail the derivation of the leading-logarithmic power correction at NLO for an arbitrary $N$-jet process.  We summarize which features of this result generalize to the NNLO level.  In Section~\ref{sect:powernnlo} we explicitly calculate the leading-logarithmic power correction at NNLO for color-singlet production mediated by both $q\bar{q}$ and $gg$ initial states.  In Section~\ref{sect:numerics} we study the numerical impact of the power corrections on the $N$-jettiness subtraction scheme.  Our main results for the power corrections at NNLO are summarized in Section~\ref{sect:powernnlo}, in the form of simple analytic expressions amenable to numerical implementation.

\section{Derivation of the NLO power correction}\label{sect:nlo}

In this section we briefly review the leading-power factorization theorem and present a detailed derivation of the leading-logarithmic power correction at NLO for an $N$-jet process.  We emphasize the features of the calculation which extend to the NNLO level.

\subsection{Review of the leading-power factorization theorem}

In the $N$-jettiness subtraction scheme~\cite{Boughezal:2015dva,Gaunt:2015pea} for a generic collider process involving $N$ jets in the final state, the $N$-jettiness event-shape observable~\cite{Stewart:2010tn}
\bea\label{eq:taudef}
\tau_N = \sum_k  {\rm min}\left[    \frac{ p_k \cdot n_a}{Q_a} , \frac{ p_k \cdot n_b }{Q_b}, \frac{ p_k \cdot n_1}{Q_1}, \dots , \frac{ p_k \cdot n_N}{Q_N} \right]\,,\quad
\eea
serves as the resolution parameter between the $N+1$ jet configuration and the doubly-unresolved limit.  Here, the $p_k$ denote the four-momenta of final-state QCD partons. 
The $n_{a,b}$ are light-like vectors for the initial beam directions and the $n_i$ are light-like vectors denoting the directions of the final-state jets in the problem. The $n_i$ are determined by pre-clustering final-state radiation using a standard jet algorithm.  The $Q_i$ are variables characterizing the hardness of the beam jets and final-state jets. The minimum in Eq.~(\ref{eq:taudef}) defines the contribution of $p_k$ to $\tau_N$ according to which direction $p_k$ is closest.  The small-$\tau_N$ cross section is derived using the all-order leading-power factorization theorem for the cross section~\cite{Stewart:2009yx}, obtained using the Soft-Collinear-Effective Theory (SCET)~\cite{Bauer:2000ew,Bauer:2000yr,Bauer:2001ct,Bauer:2001yt,Bauer:2002nz}.  We schematically write the differential cross section in the small-$\tau_N$ limit as
\bea\label{eq:LPfact}
\mathrm{d}  \sigma(\tau_N) \sim {\rm Tr} ( H \cdot {\cal S}_N ) \, \otimes \,{\cal  B}_a \, \otimes \,{\cal B}_b \, \
\prod_i^{N} \, \otimes \, {\cal J}_i , 
\eea
where the operator definition of each components can be found in Ref~\cite{Stewart:2009yx}. The beam function ${\cal B}$~\cite{Gaunt:2014xga,Gaunt:2014cfa}, the jet function ${\cal J}_i$~\cite{Becher:2006qw, Becher:2010pd} and 
the soft function ${\cal S}_N$ for jets~\cite{Boughezal:2015eha} and for the massive case~\cite{Li:2016tvb}
 are all known to the required NNLO level.

The results of~\eq{eq:LPfact} expanded to fixed order in the strong coupling constant can also be obtained 
using the method of regions~\cite{Beneke:1997zp}.  This entails expanding the full QCD matrix elements and the phase space consistently in $\tau_N$ assuming either a soft-momentum scaling $p_s \sim Q \tau_N $ or a 
collinear momentum scaling $p_c \sim Q\left(1,\sqrt{\tau_N},\tau_N\right)$\footnote{We note that there is a collinear mode of this form for both the initial beam directions and the final-state jet directions}, which exhaust the possible leading singular regions.  In writing the collinear momentum scaling we have adopted the usual Sudakov decomposition of $p_c$. In the method of regions approach, the collinear and soft behaviors are disentangled and the fixed order results of~\eq{eq:LPfact} can be recovered as the sum of soft and collinear contributions. To avoid double counting between the collinear and  soft regions, a zero-bin subtraction~\cite{Manohar:2006nz} is usually required in order to reproduce the leading singular results of QCD. 

The leading-power factorization theorem is exact only when $\tau_N \to 0$. Therefore in practical applications of $N$-jettiness subtraction a very small $\tau_N^{cut}$ is introduced, and~\eq{eq:LPfact} is used below $\tau_N^{cut}$.  The below-cut cross section receives power corrections with the leading behavior $\alpha_s^n  \log^{2n-1}(\tau_N)$.  Including the power corrections would allow larger $\tau_N^{cut}$ to be used, and could in principle improve the numerical performance of the $N$-jettiness subtraction scheme.

\subsection{Power corrections at NLO}

In this section, we calculate the leading power correction at NLO to the $N$-jettiness factorization theorem in~\eq{eq:LPfact}. 
We illustrate our derivation focusing on the $\tau_1$ measurement in deep inelastic scattering (DIS).  This general case contains both an initial-state hadron beam and a final-state jet, allowing us to demonstrate all possible technical details. We define $N$-jettiness using the hardness measures $Q_a^{-1}  =   x \sqrt{s}/ s_{ll}$ and $Q_1^{-1}  = 2E_J/s_{ll}$, where $E_J$ is the jet energy, although the derivation holds more generally.  Although we show in detail the derivation of the leading-logarithmic terms ($\log{\tau}_1$ at NLO) in the leading-logarithmic power correction (LL$_P$), the remaining terms which scale as $\tau_1$ can also be evaluated in a straightforward manner. We will present the NLO power correction for DIS, jet production in $e^+e^-$ collision and Higgs production in $gg$ fusion in the end of this section, which represent the possible cases for the $N$-jettiness NLO power correction. The NLO power corrections for a generic $N$-jet process can be obtained as a linear combination of these three cases with color factors assigned properly, plus corrections derived from the $N$-jet Born matrix element. 

\subsubsection{General features for LL$_P$ at NLO}
Before showing the explicit calculation, we note that on general grounds, the NLO cross section for measuring $\tau_N$ takes the form
\bea\label{eq:nlosig}
\frac{\mathrm{d} \sigma^{(1)}}{\mathrm{d} \tau_N} =  
\frac{1}{\tau_N} \,
\int^{1-  f(\tau_N) } \,   \frac{\mathrm{d} z }{1 - z} \, {\cal N}(\tau_N, 1-z) \,, 
\eea
where $z$ parameterizes the energy of the final-state radiation and $\tau_N$ controls the collinear singularity. 
As $z \to 1$, the emission becomes soft and forces $\tau_N \to 0$, which justifies the upper limit in the $z$-integral. Here the form of $f(\tau_N)$ depends on the 
kinematic details. $f(\tau_N)$ starts at ${\cal O}(\tau_N)$ and vanishes as $\tau_N \to 0$. The numerator ${\cal N}$  is a regular function in both limits $z\to 1$ and $\tau_N \to 1$, and includes information from both the matrix element and possible phase space cuts. If we expand ${\cal N}$ in terms of $\tau_N$ and $1-z$
\bea
{\cal N}(\tau_N,1-z)  = {\cal N}_0(0,1-z)  \, + \, {\cal N}_1(0,0) \, \tau_N +  \dots \,,
\eea
we can see that $ {\cal N}_0$ will give rise to the contribution covered by 
the leading-power factorization theorem in~\eq{eq:LPfact}. ${\cal N}_1$ will generate the $\log(\tau_N)$ power
correction, while the remaining terms will contribute to terms linear in $\tau_N$ or higher orders in powers of $\tau_N$. Therefore determining
${\cal N}_1$ in which the emission becomes soft is our primary goal. 

We begin by listing general features of the LL$_P$ that we find at NLO that hold true at NNLO as well.

\begin{itemize}

\item The results are free of divergences. All $\epsilon$-poles cancel among themselves in the power corrections.  This is a clear requirement of the LL$_P$; the differential cross section in $\tau_N$ is a physical observable free of divergences, and it can be expanded in the small-$\tau_N$ limit to obtain the LL$_P$.

\item The LL$_P$ comes solely from the soft limit.  The configurations which contribute to the leading-power expression in Eq.~(\ref{eq:LPfact}) but not to its leading logarithms, do not give rise to LL$_P$.  An example of a leading-power configuration that does not contribute to the LL$_P$ is in the $g l \to q q l $ channel, when one of the $q$ is grouped with the beam to contribute to $\tau$. 

\item Soft quarks contribute to the LL$_P$.  Their contributions can be determined unambiguously from the leading power splitting kernels.  

\item Power divergences  occur in the power corrections due to expanding terms such as $(1 - z' + z' \tau_N)^{-1}$ in terms of $\tau_N$. The power divergence can be eliminated by rescaling $z' = z/(1-\tau_N) $. The rescaling leads to the appearance of derivatives of the parton distribution functions (PDFs) $\partial_x f(x) $ in the power correction. 

\end{itemize}

We have used both a rigorous QCD calculation and the method of regions to obtain our results. We have checked that these two methods lead to identical expressions. In the following discussion we present the calculation using the method of regions, since we will later extend this approach to the NNLO level.  
A similar procedure has been applied in Ref.~\cite{Bonocore:2014wua,Bonocore:2015esa} with focus on the threshold power corrections in Drell-Yan production. 
We discuss the derivation using the DIS process as an example.
In DIS, following the $N$-jettiness definition Eq.~(\ref{eq:taudef}), the calculation can be organized by the beam and the jet contributions depending on whether the radiation is grouped with 
the beam or with the jet direction to contribute to the jettiness. We note that both the beam and jet
contributions are well-defined, IR-safe physical observables. The global $N$-jettiness is the sum 
of these two contributions. In the method of regions, both beam and jet contributions receive their dominant 
part from the configurations in which the radiation momenta become collinear (scales as collinear) or soft (scales as soft) which
will be defined later when we discuss the beam and jet contributions. In each contribution, beam or jet, the final results will be the sum of the collinear and the soft sectors, with proper zero-bin (overlap between collinear and soft) subtracted out to avoid double counting.  

\subsubsection{ Beam contribution}

We consider the real-radiation correction to the DIS process: $xP_a + p_b \to l + q + k$.   At NLO, the virtual corrections do not lead to power corrections, since their full contributions to the cross section have already been included Eq.~(\ref{eq:LPfact}).  The phase space for this process can be written as
\bea\label{eq:psbeam}
\mathrm{d}\Phi_{\mathrm{DIS}}^{(1)}& = &
\frac{1}{(2\pi)^{2d-3}} \,
\mathrm{d} x \,
\mathrm{d}^d l \, \delta(l^2) \,
\mathrm{d}^d q \, \delta(q^2) \,
\mathrm{d}^d k \, \delta(k^2) \, \nn \\
&&\times 
\delta^{(d)}( x P_a + p_b - l - q - k ) \,
\delta\left(
\tau_1 - \frac{2 x_B }{2p_b \cdot l } P_a \cdot k  \right) \,
\Theta(p_a,q,k)  \,. 
\eea
Here, the measurement function $\Theta$ enforces that
$k$ is grouped with the hadron beam to contribute to $\tau_1$ which defines the beam contribution. $P_a$ and $p_b$ are the four-momenta for the incoming hadron and lepton respectively. The Bjorken-$x$ relates
the hadron momentum $P_a$ to the  partonic momentum as $p_a = x \, P_a$. $l$ denotes the  outgoing lepton $4$-momentum, while $q$ and
$k$ are the momenta for final-state partons.  We always assume that $k$ is potentially unresolvable while $q$ is always hard. 
The last $\delta$-function involving $\tau_1$ defines the jettiness observable $\tau_1$, in which  $x_B \equiv  ( 2 p_b \cdot l )/( 2 P_a \cdot (p_b - l ))$.  We note that this definition makes $\tau_1$ dimensionless. The superscript on the differential phase space denotes that this is an NLO expression.

To proceed, 
we parameterize the momentum $k$ using the light-cone coordinates defined by $p_a^\mu$ and $q^\mu$:
\bea
k^\mu = \frac{q\cdot k }{p_a \cdot q} p_a^\mu 
+ \frac{p_a\cdot k }{p_a \cdot q} q^\mu  + k_\perp^\mu \,.
\eea
Note that in the beam contribution $p_a \cdot k \sim {\cal O}(\tau_1 Q^2)$
and $k_\perp^2 \sim q\cdot k \, p_a \cdot k$ as required by the jettiness definition and the on-shell condition for $k$.   
Here $q\cdot k$ can either be large (of order $Q^2$) or small (of order $ Q^2 \tau$), which defines the collinear scaling and soft scaling of the momentum, respectively. However for the collinear scaling, when we perform the phase space integration, the momentum component $q\cdot k$ will unavoidably reach a region in which $q\cdot k \sim Q^2 \tau$. In this region the momentum scaling overlaps with the soft scaling which defines the zero-bin. We therefore need to subtract out the zero-bin in the collinear sector to avoid double counting. 

In the following, we detail the evaluation of the  collinear scaling contribution 
$q\cdot k \sim Q^2$. The soft ones can be obtained similarly by assuming $q\cdot k \sim Q^2 \tau_1 $.
Writing $k$ in terms of the light-cone decomposition, the $\delta$-function for  energy-momentum conservation in Eq.~(\ref{eq:psbeam}) can expressed as
\bea\label{eq:momcon}
\delta^{(d)}\left(
\left[ 1 - \frac{2q\cdot k}{s_{ll}/z'}     \right] x P_a 
+ p_b - l 
- \left[ 
1 + \frac{2p_a \cdot k }{ s_{ll}/z'}
\right]q 
\right) \,,\quad \quad
\eea
where we have introduced a variable  
$
 z' = \frac{s_{ll}}{2p_a \cdot q}  
$, 
with $s_{ll} = 2 p_b \cdot l$. 
The collinear scaling $q\cdot k \sim Q^2$ means $1-z' \sim {\cal O}(1)$. 
Here we have dropped in the $\delta$-function the $k_\perp$ dependence. 
This is allowed to the logarithmic accuracy in which we are interested, since any term linear in $k_\perp$ will vanish 
after being averaged over the solid angle of $k_\perp^\mu$, while any $k_\perp^2$ will scale as $\tau_1 \, (1-z') $ which will
not contribute to the logarithms in the power correction.~\footnote{Dropping $k_\perp$ will affect the  value of $2q\cdot k$ but it has no overall effect on the final logarithmic power correction.} 

From the $\tau_1$ definition in~\eq{eq:psbeam} and the momentum conservation expression in~\eq{eq:momcon}, it is
straightforward to find
\bea
2 p_a \cdot k &=& \frac{s_{ll}}{z} \, \tau_1 \,, \quad
2 q \cdot k = \frac{s_{ll}}{z} \Big( 1 - z  \Big) \Big(1-\tau_1 \Big) \,, \nn \\
2 p_a \cdot q &=& \frac{s_{ll}}{z} \Big(1-\tau_1 \Big) \,, 
\eea
To avoid the possible occurrence of power divergences in deriving the power correction as a consequence 
of expanding $(1 - z' + z' \tau_1)^{-1}$ in $\tau_1$,
we have rescaled the variable $z'$ using
\bea
z'  = \frac{ z }{1-\tau_1} \,,
\eea
This fixes the ambiguity of order $\tau_1$ in defining ${\cal O}(1) $ variables. We note that the upper bound
of the $z$-integration is $1-\tau_1$. However, we can safely integrate $z$ all the way to $1$.  The range $[1-\tau_1,1]$ is where $q\cdot k \sim Q^2 \tau_1$ and the zero-bin subtraction should be performed as we discussed before. After correctly implementing the zero-bin subtraction, this range receives no LL$_P$.  

The phase space in~\eq{eq:psbeam} is factorized into a Born piece which only involves leading power momenta which scale as ${\cal O}(Q)$, and a radiative phase space:
\bea
\mathrm{d}\Phi_{\mathrm{DIS}}^{(1)} = \mathrm{d} \Phi_{\mathrm{Born}} \times \mathrm{d} \Phi_{R} .
\eea
The Born phase space is
\bea\label{eq:BORNPS}
 \mathrm{d} \Phi_{\mathrm{Born}} &=& \mathrm{d} x_{ B}
\frac{\mathrm{d}^d l   \delta(l^2) \, }{(2\pi)^{d-1}    } 
\frac{\mathrm{d}^d q_{ B}   \delta(q_{ B}^2) }{(2\pi)^{d-1} }   
\,  (2\pi)^{d} \delta^{(d)}( x_{ B} P_a + p_b -l - q_{ B}) \,, 
 \eea
with
\bea
x_B = z \, x \,, \quad \quad 
q^\mu_B = \frac{1}{1-\tau_1} q^\mu .
\eea
The radiative phase space is
\bea\label{eq:PSR}
\mathrm{d} \Phi_{R}  &= & \frac{(1-\tau_1)^{d-3}}{z} 
\frac{ \mathrm{d}^d k \, \delta(k^2) \,  }{(2\pi)^{d-1}} \, 
s_{ll} \, (1-\tau_1) \nn \\ 
&& \times \frac{ \mathrm{d} z }{z } \,
  \delta\left( \frac{s_{ll} (1-z) }{z} (1-\tau_1) - 2 q\cdot k     \right) \,
\delta\left(
\tau_1 - \frac{2z }{s_{ll} } p_a \cdot k  \right) \,
\Theta(p_a,q,k) \,,
\eea
where the first factor $(1-\tau_1)^{d-3}/z$ is the Jacobian ${\cal J}(\tau_1,z) $ due to the variable change from $\{x\,,q^\mu\}$ to $\{x_B\,,q^\mu_B   \}$ needed to obtain the Born phase space. 
Assuming the collinear scaling $ q\cdot k \sim Q^2 $, $p_a \cdot k \sim Q^2 \tau_1 $ and $z \sim 1$, we find 
the phase space for the collinear sector in the method of regions approach 
\bea\label{eq:PScollb}
\mathrm{d} \Phi^c_{R}  =   {\cal J}
\frac{  s_{ll}  \,  \Omega_{2-2\epsilon} }{4 (2\pi)^{3-2\epsilon}}   \frac{ \mathrm{d} z }{z} 
\left(   \frac{ s_{ll} (1-z) }{z} \tau_1 \right )^{-\epsilon}
\Theta(p_a,q,k) \,, \quad\quad
\eea
where $\Omega_{2-2\epsilon}$ is the $2-2\epsilon$ dimensional solid angle from $k_\perp$. 

The phase space for the soft sector and the zero bin can be derived
straightforwardly from~\eq{eq:PSR} by assuming $q\cdot k \sim Q^2 \tau_1$ and dropping the $q\cdot k $ dependence in the fist $\delta$-function.  The explicit form will not be shown here. We also note that the measurement function $\Theta$ can be set to 1 in the collinear sector up to the accuracy we are working with.

The contribution from expanding the PDFs in the beam sector is 
\bea
f_i(x) = f_i\left( \frac{x_B}{z} \right) \,,
\eea
 which receives no power correction.  There is also  contribution from the flux
\bea
\frac{1}{2 x P_a \cdot p_b } = \frac{1}{ 2x_B P_a \cdot p_b } \, z \,,
\eea
which again receives no power correction in $\tau_1$.   Before we turn to the matrix element evaluation, we notice that we can make the simplification
\bea
&& (2 l \cdot q)^2  
\sim  \left(      2 l \cdot q_B   \right)^2 
 -  2  \tau_1 \, \left( 2 l \cdot q_B                \right)^2  \,, \nn \\
&& ( 2 p_b \cdot q  )^2  \sim \left(    2 p_b \cdot q_B \right)^2 
- 2 \tau_1 \, \left( 2 p_b \cdot q_B      \right)^2 \,,
 \eea
where we have omitted terms of order ${\cal O}(\tau^2)$.

We now consider the squared matrix elements for this process assuming the same collinear scaling as in the phase space.  There are two partonic channels to consider, $ql \to l^{\prime}q^{\prime}g$ and $gl \to l^{\prime}q\bar{q}$.   
We first note that in the beam sector the $gl$ channel does not contribute to the LL$_P$,  as discussed in our presentation of the general features of the NLO LL$_P$.
For the $ql$ channel, 
using the simplification of scalar products above we find that the matrix element for the $ql$ channel with a soft gluon reduces to
\bea\label{eq:Mqlbeam} 
 |{\cal M}_{ql}^{(1)}|^2 \to \frac{1}{s_{ll}} \, |{\cal M}^{(0)}|^2 \, 
( 8\pi\alpha_s C_F ) \, \frac{2   }{\tau_1 \, ( 1- z ) }  
\,,
\eea
and for ql channel with an unresolved quark we have the matrix element 
\bea\label{eq:Mglbeam}
 |{\cal M}_{ql}^{(1)}|^2 \to \frac{1}{s_{ll}} |{\cal M}^{(0)}|^2 \, (8\pi\alpha_s C_F) \frac{1}{\tau_1 (1-z)} \, \tau_1 .
\eea
${\cal M}^{(0)}$ is the LO matrix element for $ql \to q'l' $ with leading-power kinematics. 
For the unresolved quark, the LL$_P$ comes
from the configuration in which a quark is emitted from the final state $q$ but grouped with the beam to contribute to $\tau_1$.
Although this ``anti-collinear-grouping" configuration 
does not contribute to the leading-power singular terms in~\eq{eq:LPfact}, it does have an effect in the power correction. We further note that to get Eq.~(\ref{eq:Mglbeam}), instead of expanding the full NLO matrix element for real emission, Eq.~(\ref{eq:Mglbeam}) can be determined directly from $s^{-1}_{gq}P_{qg}(x) $ by relating $s_{qg} \sim (1-z)$ and $1- x \sim \tau_1$.

Combining the phase space in~\eq{eq:PScollb} and the matrix elements in~\eq{eq:Mqlbeam} and~\eq{eq:Mglbeam}, we find the power correction from the
collinear sector to be
\bea \frac{1}{\epsilon}
 \frac{  \alpha_s C_F }{   \pi  } \, \left( 
\frac{1}{2 } 
\right) 
 \, \left(   \frac{\tau_1  s_{ll}}{\mu^2}  \right)^{-\epsilon} \,
 \,
\mathrm{d} \Phi_{\rm B}
|{\cal M}_0|^2 \,,
\eea
where from the effective theory point of view, $\sqrt{ \tau  s_{ll} }$ fixes the collinear scale, as expected. 

Following similar procedure by assuming $q\cdot k \sim \tau Q^2$, we find the contribution from the soft scaling is 
\bea 
- \frac{1}{\epsilon}
 \frac{  \alpha_s C_F }{   \pi  } \, \left( 
\frac{1}{2 } 
\right) 
 \, \left(   \frac{\tau_1^2  s_{ll}}{\mu^2}  \right)^{-\epsilon} \,
 \,
\mathrm{d} \Phi_{\rm B}
|{\cal M}_0|^2 \,,
\eea
Once we combine these, we arrive at the following logarithmic power
correction from the beam contribution:
\bea  
\mathrm{d} {\hat \sigma}_{\mathrm{beam}}^{(1)} = - \mathrm{d} {\hat \sigma}^{(0)}
\frac{\alpha_s C_F }{\pi}
\left( 
\frac{1}{2} 
\right) 
 \,  f_q(x_{B}) \,
L   \,,
\eea
where $\mathrm{d} \hat{\sigma}_0 = \mathrm{d} \Phi_B |{\cal M}_0|^2$ and 
$L = \log\left( \frac{\tau_1 s_{ll}}{\tau_1^2 s_{ll}}  \right) $.
The ratio in the logarithm reflects the scale hierarchy between the collinear and soft sectors.  
In the power corrections, no remaining singularities in $\epsilon$ should arise. 
All the $\epsilon$ poles must cancel amongst soft and collinear regions since the beam contribution itself is a well defined physical observable.  We indeed find that they do. 
 
\subsubsection{ Jet contribution}
We next study the jet contribution following the same steps as for the beam region.  Momentum conservation for $xP_a + p_b \to l + q + k$ can be simplified to
\bea\label{eq:momconfin}
\delta^{(d)}\left(
\left[ 1 - \frac{2q\cdot k}{z' \, s_{ll} }     \right] x P_a 
+ p_b - l 
- \left[ 
1 + \frac{2p_a \cdot k }{z' \, s_{ll} }
\right]q 
\right) \,. \quad \quad
\eea
Unlike the beam contribution, we have defined $z' = \frac{2p_a \cdot q}{s_{ll}}$. The one-jettiness definition becomes 
\bea
\tau_1 =  \frac{2}{s_{ll}}  \, \left( 1+ \frac{2k\cdot p_a}{z' \, s_{ll}} \right) q\cdot k    \,,
\eea
Deviations from this result due to pre-clustering the partons with different infra-red safe jet algorithms will not contribute to the LL$_P$.  We can parameterize 
\bea
&&
2p_a \cdot q = z \, s_{ll} \,(1+\tau_1) \,, \nn \\
&&
2 k\cdot q = z \, s_{ll} \, \tau_1 \,, \quad 
2 k \cdot p_a = z \, s_{ll} \, (1+\tau_1) \, \frac{1-z}{z} \,,
\eea
where as required by the jettiness measurement, $k\cdot q \sim Q^2 \tau_1$ while $k\cdot p_a$ can either be
of order $Q^2$ or $Q^2 \tau_1$, which defines the collinear scaling and soft scaling, respectively. In the following, we will focus 
on the collinear sector in which $k\cdot p_a \sim Q^2$. This also implies that $1-z \sim {\cal O}(1)$.  
Here we have rescaled $z'$ using 
\bea
z' = z(1+\tau_1) \,,
\eea
to avoid power divergence from expanding the matrix element in $\tau_1$.

Similar to the beam contribution, the phase space is factorized into a Born part
\bea\label{eq:BornJET}
\mathrm{d} \Phi_{\mathrm{Born}} &=& \mathrm{d} x_{ B}
\frac{\mathrm{d}^d l   \delta(l^2) \, }{(2\pi)^{d-1}    } 
\frac{\mathrm{d}^d q_{ B}   \delta(q_{ B}^2) }{(2\pi)^{d-1} }   \, 
(2\pi)^{d} \delta^{(d)}( x_{ B} P_a + p_b -l - q_{ B}) ,  
\eea
with
\bea
x_B = \frac{1}{1+\tau_1} \, x \,, \quad \quad
q_B^\mu = \frac{1}{z} \, q^\mu \,,
\eea
and a radiation piece 
 \bea
 \mathrm{d} \Phi_R &=& 
(1+\tau_1) z^{d-3}\,
\frac{   \mathrm{d}^d k    \delta(k^2) \,  }{(2\pi)^{d-1}} 
\,
s_{ll}(1+\tau_1) \,  \nonumber \\ && \times  \mathrm{d}z  \, 
\delta\Big(
s_{ll} (1-z)(1+\tau_1) - 2 p_a \cdot k  
\Big) \,
\delta\left(
\tau_1 - \frac{2 k \cdot q}{z s_{ll}}
\right)
 \,.
\eea 
The first factor is again the Jacobian ${\cal J}=(1+\tau_1) z^{d-3}$ from the variable change
$\{x, q\}  \to \{x_B,q_B\}$ needed to reach the Born phase space.  Assuming that $k^\mu$ follows the collinear scaling $ p_a \cdot k \sim Q^2 $, $q \cdot k \sim Q^2 \tau_1 $ and $z \sim 1$, we find 
the phase space for the collinear sector in the method of regions:  
\bea
\mathrm{d} \Phi_R = 
{\cal J}
\,
\frac{s_{ll}  \, \Omega}{4 (2\pi)^{3-2\epsilon}}\,
\mathrm{d} z \, 
\Big(
 \,  s_{ll} (1-z)  \,  \, \tau_1 
\Big)^{-\epsilon} \,.
\eea
Deriving the phase space for the soft and the zero-bin subtraction follows similar steps but with the additional assumptions that $p_a \cdot k  \sim Q^2 \tau_1 $.  The power correction due to expanding the PDFs around $x_B$ is given by
\bea\label{eq:rescalepdf}
f_i(x) 
=  f_i(x_B) +    \tau_1 \, \Big[ x \, \partial_x \, f_i(x) \Big] | _{x=x_B} \,  \,,
\eea
while the flux contributes as
\bea
\frac{1}{2 x P_a \cdot p_b} = \frac{1}{2 x_B P_a \cdot p_b}  \frac{1}{1+\tau_1} \,.
\eea

Having derived the power correction coming from the phase space, we move onto the matrix elements.  Noting the simplification 
\bea
&&(2p_a \cdot p_b)^2 \sim (1+2 \tau_1 ) \left(  2p_{a,B} \cdot p_b  \right)^2  \,, \nn \\
&&(2p_a \cdot l )^2 \sim (1+2\tau_1 ) \left(   2p_{a,B} \cdot l \right)^2 \,,
\eea
in which all terms linear in $k_\perp$ or proportional to $\tau_1(1 - z)$ have been dropped,
we find that the matrix element for the $ql$ channel with an unresolved gluon reduces to 
\bea
|{\cal M}_{ql}^{(1)}|^2 \to \frac{1}{s_{ll}} \, |{\cal M}_0|^2 \,
(8\pi\alpha_s C_F) \, 
\frac{1}{z} \,
\frac{2 }{\tau_1 (1-z)} \,,
\eea
and the matrix element for the $gl$ channel with an unresolved quark simplifies to
\bea
|{\cal M}_{gl}^{(1)}|^2 \to \frac{1}{s_{ll}} \, |{\cal M}_0|^2 \, (8\pi\alpha_s C_F) \frac{1}{1-z} \,,
\eea
This second matrix element comes from a soft quark  emitted from the initial state $p_a$ but grouped with $q$ to contribute to $\tau_1$.
Again ${\cal M}_0$ is the LO matrix element for $ql \to q'l' $ with leading-power kinematics.

The contribution from soft sector can be obtained in a similar manner with the assumption that $p_a \cdot k \sim Q^2\tau_1$. Putting together all components, we find the power correction from jet contribution is
\bea
\mathrm{d} {\hat \sigma}_{\mathrm{jet}}^{(1)} = \mathrm{d}  \hat{\sigma}_0
\frac{\alpha_s }{\pi}
\left( 
C_F  
  \Big[  x \partial_x f_q  \Big] 
+
\frac{T_R }{2} \sum_{i = - N_F}^{N_F}  Q_i^2  \, f_g
  \right)  
L
\,. \quad
\eea
Again, the $\epsilon$-poles are absent in the final result, since the jet contribution is itself a physical observable. 
Here the PDFs are evaluated at $ x = x_B $. In the $gl$ channel, we have normalized the result to the $ql$ channel color and spin average by multiplying by a factor of $\frac{N_C}{N_C^2 - 1}$.  $\mathrm{d}  \hat{\sigma}_0$ is the Born-level matrix element and phase space with the PDF removed.

\subsection{Summary of NLO results}

\subsubsection{NLO power correction for DIS}

Combining the contributions from the beam and jet contributions found in the previous sections, we obtain the full power correction the DIS $\tau_1$ distribution as
\bea\label{eq:NLOpowerdis}
\mathrm{d} {\hat \sigma}_{\mathrm{DIS}}^{(1)} = \mathrm{d} {\hat \sigma}_0 \, 
\frac{\alpha_s}{2 \pi}
\left(  C_F \, (-  1 + 2 \, x \, \partial_x)
 \,  f_q  
 \, 
  +
 T_R \sum_{i = - N_F}^{N_F} \, Q_i^2  \, f_g
  \right) 
L \,,  
\eea
where $Q_i$ is the electric charge of the $i$th quark, $L = - \log(\tau_1) $ and the PDFs are evaluated at $ x = x_B $.  We note that the power correction for DIS comes from the Jacobian ${\cal J}$ in~\eq{eq:PScollb},  
from expanding PDFs in~\eq{eq:rescalepdf}, and from the soft quark configuration. 

\subsubsection{NLO power correction for hadronic production in $e^+e^-$ collisions}

The power correction for $N$-jettiness (or equivalently thrust) for hadronic production in $e^+e^-$ collisions is found to be
\bea\label{eq:NLOpoweree}
\mathrm{d} {\hat \sigma}_{e^+e^- \to \mathrm{hadrons}}^{(1)} =  \mathrm{d} {\hat \sigma}_0 \, 
\frac{\alpha_s}{2 \pi} C_F 
\left( \, -  2 
   \right) \,
L \,,  
\eea
where $L = - \log(\tau)$.  The derivation follows closely the one presented for the DIS beam contribution. 
The result reproduces the known logarithmic power correction found from both fixed-order calculations~\cite{Kramer:1986mc} and within the framework
of SCET~\cite{Freedman:2013vya}, which demonstrates the validity of the method of regions approach in studying the power corrections. Here the power correction comes from Jacobians due to rescaling the Born momentum $q_1$ or $q_2$ to $q_{1,{\rm B}}$ or $q_{2,{\rm B}}$ in the process
$l^+ l^- \to q(q_1) {\bar q}(q_2) g(k) $, and from the soft-quark matrix element.

\subsubsection{NLO power correction for $ggH$ and Drell-Yan}

Following a similar procedure as in the previous sections, we can evaluate the power corrections for $ggH$. We define $\tau_0$ using the hardness measures $Q_a = Q_b = 1$ in~\eq{eq:taudef}, implying that $\tau_0$ has units of energy.  For the $gg$ channel, we find
\bea\label{eq:NLOpowerggH}
\mathrm{d} {\hat \sigma}_{gg\to H}^{(1)} =  \frac{\alpha_s C_A}{2 \pi} \mathrm{d} {\hat \sigma}_0  \bigg( L \,
\left[
2 \, x_1  \, \partial_{x_1} \,  
\right]   {\cal L}_{gg}  \,
+ \{x_1 \leftrightarrow x_2,  Y \leftrightarrow -Y \}\bigg) \,,
\eea
where $L =\frac{e^{Y}}{m_H} \,  \log \left(   \frac{\tau_0 m_H e^Y}{\tau_0^2}   \right)$ and ${\cal L}_{gg} = {\cal L}_{gg}(x_1,x_2)$ is the gluon-gluon luminosity.  We note that $\mathrm{d} {\hat \sigma}_0$ is the Born-level differential cross section for $gg \to H$ with the PDFs removed, and $Y$ is the rapidity of  the Higgs.  For the $qg + gq$ channel, we have
\bea\label{eq:NLOpowerggHq} 
\mathrm{d} {\hat \sigma}_{gq+qg\to H}^{(1)} = \frac{\alpha_s C_F}{2\pi}  \mathrm{d} {\hat \sigma}_0 \bigg( L \, {\cal L}_{g_1 q_2}   
 +  \{x_1 \leftrightarrow x_2,  Y \leftrightarrow -Y \} \bigg)\,.
\eea
Here, $L$ follows the definition above and ${\cal L}_{g_1q_2}$ is the gluon-quark luminosity.

For Drell-Yan production of lepton pairs through a vector boson $V$, the $q^i{\bar q}^j$ channel power correction gives
\bea\label{eq:LPqq}
\mathrm{d} {\hat \sigma}_{q\bar{q}\to V}^{(1)} = \frac{\alpha_s C_F}{2 \pi}\mathrm{d} {\hat \sigma}_0 \bigg( L \,
\left[
2 \, x_1  \, \partial_{x_1} \,  
\right]   {\cal L}_{q^i{\bar q}^j}  
+ \{x_1 \leftrightarrow x_2,  Y \leftrightarrow -Y \} \bigg)\,,
\eea
while the $qg + gq$ channels contribute
\bea
\mathrm{d} {\hat \sigma}_{gq+qg\to V}^{(1)} = \mathrm{d} {\hat \sigma}_0 \bigg(  \sum_{j=-N_F}^{N_F}Q_j^2 \,  V_{ji}
\, \frac{\alpha_s T_R}{2\pi}  L \, {\cal L}_{q^i_1 g_2}  
+  \{x_1 \leftrightarrow x_2,  Y \leftrightarrow -Y \} \bigg)\,,
\eea
where $Q_j$ is the change carried by the final-state soft quark $j$. $V_{ji} = \delta_{ji}$  for $Z$ production and is the CKM matrix for $W$ production. Since we are not measuring the final state flavors a sum over $j$ arises. $\mathrm{d} {\hat \sigma}_0$ is again the Born-level cross section for $q\bar{q} \to V$ with the PDFs removed.  One should also replace $m_H \to m_V$ in the definition of the logarithm $L$ for the Drell-Yan case.  In both $ggH $ and Drell-Yan cases, the net NLO power correction comes from expanding the PDFs and from the soft quark matrix element.

\section{Derivation of the NNLO power correction }\label{sect:powernnlo}

We next present the derivation of the power corrections at NNLO. At ${\cal O}(\alpha_s^2)$, we have to deal with both real-virtual (RV)
and real-real (RR) contributions. The double-virtual correction has been entirely included in the leading power factorization theorem in
Eq.~(\ref{eq:LPfact}). For RV, the phase space integration is identical to the NLO phase space, For RR, the calculation is more involved.  We begin by stating some of the features of the LL$_P$ that we observe at the NNLO level. 
\begin{itemize}

\item All $\epsilon$-poles cancel between RV and RR in the LL$_P$.

\item Soft limits lead to LL$_P$ at NNLO just like at NLO. The LL$_P$ soft currents for the $qg$ channels for both Higgs and Drell-Yan production are deducible from the leading-power splitting kernels. 

\item Explicit calculations show that the LL$_P$ comes from the strongly-ordered limit of the matrix elements at NNLO. For instance, in the case of a $qg$ final state, $E_g \ll E_q$. We also notice from our explicit calculations  that 
within the strongly-ordered limit, 
the final results for the LL$_P$ can be obtained using an independent-emission approximation in the phase space integrals.  

\item Configurations (the Abelian piece) which contribute to the leading logarithms in the leading-power result also contribute to the LL$_P$.  Their contribution to the LL$_P$ at NNLO can be written as a convolution in $\tau_N$ between an NLO leading-power leading logarithmic contribution and an NLO LL$_P$. 

\end{itemize}
Although we can not prove or disprove on general grounds these observations seen in our explicit calculations, we conjecture that all of the above features observed for zero-jettiness in color-singlet production generalize to a generic $N$-jet case. 

At NNLO, the full calculation is more lengthy than at NLO.  Here we sketch the procedure of obtaining the final result, making sure to discuss all relevant features.  We consider the following examples from the Drell-Yan process to illustrate our method: the RV corrections coming from the $qg \to Vq$ channel, and the RR corrections to the $qg \to Vqg$ channel.

\subsubsection{Sketch of the RV calculation}

We start with RV.  The construction of the phase space follows identically the procedure at NLO.  The only complication is that the matrix elements no longer have a Taylor series in $\tau_0$.  They possess fractional powers coming from the loop integrations that must be isolated in order to obtain the correct logarithms, since the logarithms come from the expansion of $\tau_0^{-n\epsilon}$ factors hitting $1/\epsilon$ poles, where $n$ is an integer.  This can be done because the soft-quark limit contributing to the LL$_P$ comes completely from the ``anti-collinear" grouping of the quark, just as at NLO and as discussed below Eq.~(\ref{eq:Mglbeam}).  The factorization of the RV matrix elements in this collinear limit are well known~\cite{Kosower:1999rx}, and contain the necessary decomposition into fractional powers. 
We then use the factorization of the matrix element in the collinear limit to write the matrix element as a sum of
\bea
{\cal M}^\ast_0 \,  {\cal M}^{(2)}_{RV,H-loop}  + c.c. &=&  - 
\frac{\alpha_s^2 \, C_F^2}{2\pi \, \epsilon^2}  \,  
\frac{16\pi  \mu^{2\epsilon} e^Y}{  k^-  \, m }  
\left( \frac{\mu}{m} \right)^{2\epsilon}  \,  |{\cal M}_0|^2
 \,, 
\eea
and 
\bea
{\cal M}^\ast_0 \, {\cal M}^{(2)}_{RV,C-loop}  + c.c. &=& 
4  \alpha^2_s
\, \mu^{2\epsilon}
 \, C_F \,  \frac{e^{Y}}{ m   }
\left[-
\frac{N_C}{\epsilon^2} \, 
+ \frac{1}{ N_C \, \epsilon^2} \left( 
\left[ \frac{\tau_0 e^Y}{m} \right]^{-\epsilon}
-1
\right) 
\right] \, \nn \\
&&\quad \quad \times \left(
\frac{ m e^{-Y}  }{\mu}
\right)^{-\epsilon}
\left(\frac{k^-}{\mu}\right)^{-1 - \epsilon} \frac{1}{\mu}  \, 
 |{\cal M}_0|^2 \,. 
\eea
In both ${\cal M}^{(2)}_{RV,H-loop} $ and ${\cal M}^{(2)}_{RV,C-loop}$, we have kept only terms that will contribute to the LL$_P$. 
We have assumed that the quark is emitted from leg $P_b$ for $P_a P_b \to V + X$ and is grouped with $P_a$ in order to contribute to the LL$_P$. Here $m$ is the virtuality of the vector boson $V$ and we use the notation that $k^- = n_b \cdot k$ and $\tau = n_a \cdot k = k^+ $.  We note that the collinear factorization of the one-loop RV amplitude schematically contains two pieces: a tree-level splitting amplitude times a one-loop amplitude for the process $q\bar{q} \to V$, and a one-loop splitting amplitude times a tree amplitude for $q\bar{q} \to V$.  The ${\cal M}^{(2)}_{RV,H-loop}$ amplitude comes from this first structure, while ${\cal M}^{(2)}_{RV,C-loop}$ comes from the second structure.  Now following the same procedure for the NLO calculations, we can get the RV contribution to the LL$_P$ straightforwardly. The RV contributions are given by the sum of 
\bea
I^{(2)}_{H-loop} = 
\left( \frac{\alpha_s}{2\pi} \right)^2\,C_F T_R \,
\frac{2}{\epsilon^3}
\frac{ e^Y}{m}  \,  
 \left[
\left( \frac{m\,e^Y\, \tau_0}{\mu^2 } \right)^{-\epsilon} 
-
\left( \frac{\tau_0^2}{\mu^2} \right)^{-\epsilon} 
\right] \left(
\frac{\mu^2}{m^2} 
\right)^{\epsilon}
\,,   
\eea
and
\bea
I^{(2)}_{C-loop} &=& 
\left(
\frac{\alpha_s}{2\pi} 
\right)^2\,
T_R  \,  \frac{e^{Y}}{ m   }
\left(
\frac{ m e^{-Y}  \tau_0}{\mu^2}
\right)^{-\epsilon}
\, \nn \\
&& \times \frac{1}{2\epsilon^3}
\left[
 \, N_C
+   \left(1- 
\left[ \frac{\tau_0 e^Y}{m} \right]^{-\epsilon}
\right) \frac{1}{N_C}
\right] \, 
 \,
\left[ 
\left( 
\frac{me^Y}{\mu} 
\right)^{-2\epsilon} 
- 
\left(
\frac{\tau_0}{\mu}
\right)^{-2\epsilon}
\right]\,.
\eea
Here we have normalized the result to the $q{\bar q}$ channel color and spin average by multiplying with a factor 
$\frac{N_C}{N_C^2-1}$, which turns an overall factor $C_F$ to $T_R$. We have 
suppressed the dependence on the luminosity and Born cross section for simplicity.  The full RV contribution to the LL$_P$ cross section for final state $qg$ is thus given by
\bea
\mathrm{d} \sigma^{(2)}_{RV,qg + gq \to V}  = \mathrm{d}\hat{\sigma}_0 \bigg( I_{H-loop}^{(2)} + I_{C-loop}^{(2)} \bigg){\cal L}_{qg}  
+ \{Y\leftrightarrow -Y, x_1 \leftrightarrow x_2 \}  \,. 
\eea

\subsubsection{Sketch of the RR calculation}

For the RR contribution we find that the matrix elements leading to the LL$_P$ can be written as a sum of  Eq.~(\ref{eq:A1}) to Eqs.~(\ref{eq:A5}), which can be derived using the soft limit of the $s_{qgg}^{-2} P_{qgg}$ splitting kernel~\cite{Catani:1999ss}. To evaluate the cross sections, both a rigorous QCD phase space integration and 
the method of regions are used to derive the final results. For RR we need to consider collinear-collinear, soft-soft and collinear-soft scalings in the method of regions, together with suitable subtraction of zero-bins. We find that the final results can be obtained using an independent emission approximation in the phase space integrals following a strongly ordered limit in which $E_g \ll E_q$.  The relevant strongly-ordered soft current can be found in Eq.~(\ref{eq:A6}).  
We organize our calculation by partitions in which both $qg$ are grouped with $n_{a,b}$ to contribute to $\tau$, and
$qg$ grouped separately with $n_a$ or $n_b$ to contribute.  We sum all the partitions to find the full contribution. Following these steps leads us to the final results.  As an example of the integration of one of the contributions in the Appendix over the relevant phase space $ \Phi[k_1,k_2]$ of $k_1$ and $k_2$, we find for
Eq.~(\ref{eq:A2}) for $q(x_a P_a) g(x_b P_b) \to V+q(k_1) g(k_2) $:
\bea
\int \mathrm{d} \Phi[k_1,k_2] \, S^{(2)}_2 =  
- \left( 
\frac{\alpha_s }{2\pi}
\right)^2 C_F T_R \, \frac{e^Y}{m} \,
\frac{1}{2 \epsilon^3} \,  
\left[
\left( \frac{me^Y \tau_0}{\mu^2}
\right)^{-2\epsilon}  
     - 
\left( \frac{me^Y \tau_0}{\mu^2}
\right)^{-\epsilon}
\left( \frac{ \tau_0}{\mu}
\right)^{-2\epsilon} 
\right]  + \dots  \,, \quad \quad
\eea
where the ellipsis denotes omitted terms which will not contribute to the LL$_P$.
 We have normalized the result to the $q{\bar q}$ channel color and spin average which turns a factor of $C_F$ to $T_R$, and have
 suppressed the dependence on the Born cross section and the luminosity. Other contributions from Eq.~(\ref{eq:A1}), Eq.~(\ref{eq:A3}), Eq.~(\ref{eq:A4}) and Eq.~(\ref{eq:A5}) are obtained similarly, to find at the LL$_P$ accuracy:
 \bea
 \mathrm{d} \sigma^{(2)}_{RR,qg + gq \to V}  = \mathrm{d}\hat{\sigma}_0
 \bigg( I_{a1}^{(2)} + I_{b1}^{(2)} + I_{ab}^{(2)} \bigg){\cal L}_{qg}  
 + \{Y\leftrightarrow -Y, x_1 \leftrightarrow x_2 \}  \,. 
 \eea
Here, the $I_{ij}^{(2)}$ are presented for completeness in Eq.~(\ref{eq:Iij}). 

\subsubsection{NNLO power corrections for $ggH$ and Drell-Yan}

We now summarize our final expressions for the LL$_P$ power corrections at NNLO.  For the $q{\bar q}$ channel in Drell-Yan we have the power correction 
\bea
\mathrm{d} {\hat \sigma}_{q\bar{q}\to V}^{(2)}  
&= & \frac{1}{2} \,  \mathrm{d} {\hat \sigma}_0 \, \left(
\frac{\alpha_s C_F}{2\pi} \right)^2
 \left[  8  \log^2  \left( \frac{ \tau_0}{\mu}   \right)   \,   - 2  
  \log^2\left( \frac{ \tau_0 m e^{Y} }{\mu^2 } \right)
      - 2  \log^2 \left( \frac{\tau_0 m e^{-Y} }{\mu^2}  \right) 
    \right] \nn \\
 &&   \hspace{-3.ex} \times 
 \bigg[  \frac{e^Y}{m} \log\left( \frac{\tau_0 }{me^Y} \right) \,
\left[
2 \, x_1  \, \partial_{x_1} \,  
\right]   {\cal L}_{q{\bar q}}     
+ \{x_1 \leftrightarrow x_2,  Y \leftrightarrow -Y \} \bigg]\,,   
\eea
which is accurate to the LL$_P$.   We note that the results can be obtained by a convolution in $\tau_0$ between
an NLO leading-logarithmic leading-power term (given in Eq.~(\ref{eq:LLLP}) for completeness) with an NLO leading-logarithmic term at subleading power, given in Eq.~(\ref{eq:LPqq}).  We note that these expressions contain sub-leading logarithms associated with the scale dependence that come ``for free" as a result of our derivation.  We keep these in our final results, although a strict expansion keeping only $\mathrm{log}^3(\tau_0)$ structures is possible.

For the $qg + gq$ channel, the RR contribution to the LL$_P$ can be derived using the soft currents in Eq.~(\ref{eq:A1}) to~(\ref{eq:A5}).  Though each term gives a somewhat lengthy expression, when summed up the final results for the power correction are simple and compact. We have for the full result, after combining RV and RR:
\bea 
\mathrm{d} {\hat \sigma}_{gq+qg\to V}^{(2)} &=& \mathrm{d} {\hat \sigma}_0 \left( \frac{\alpha_s}{2\pi} \right)^2 \,
\Big( C_F  + C_A  \Big)  \, T_R \nn \\  
 & & \hspace{-9.ex} \times \bigg(
 \frac{e^Y}{m} \sum_{j} \, Q_j^2 \, V_{ji} \, {\cal L}_{q_i g} 
  \log^2 \left(  \frac{ m e^Y }{\tau_0 }       \right)
\log \left(
\frac{    \tau_0 e^Y    }{ m } 
\right)  +  \left\{ Y \leftrightarrow - Y\,, x_1 \leftrightarrow x_2  \right\} \bigg)  .
\eea
Here we have again showed the explicit dependence on the quark charge $Q_j$, and the CKM matrix $V_{ji}$ in the case of $W$-boson production (this can be set to a Kronecker delta in the case of $Z$-boson or $\gamma^{*}$ production).
We note that the results for the RR contribution can be obtained using Eq.~(\ref{eq:A6}), which is the strongly ordered limit of Eq.~(\ref{eq:A1}) to~(\ref{eq:A5}) in which $E_g \ll E_q$.  In the strongly-ordered limit, the matrix element factorizes into a product of a sub-leading soft quark current and a known leading-power soft-gluon current involving three eikonal directions. We note that  $\mathrm{d} {\hat \sigma}_0$ is the Born-level differential cross section for the $q\bar{q} \to V$ process with the PDFs extracted.

For gluon-fusion Higgs production, the LL$_P$ contributions can be obtained in the same way as for Drell-Yan. For the Abelian case, it is found that the contributions can again be obtained by convoluting the NLO leading-logarithmic terms at leading power with the NLO LL$_P$. For the $qg$ final state, the RV can be extracted from, for instance, the $1$-loop correction to $s^{-1}_{q{\bar q}}P_{g \to q{\bar q}}$~\cite{Kosower:1999rx} with the proper crossings. The RR can be derived by suitably changing the color factors in RR for Drell-Yan. After performing the phase space integrations, we find that all the $\epsilon$-poles cancel as required, and the LL$_P$ for the $ggH$ are given by:  
\bea
\mathrm{d} {\hat \sigma}_{gg\to H}^{(2)}  
&= & \frac{1}{2} \,  \mathrm{d} {\hat \sigma}_0 \, \left(
\frac{\alpha_s C_A}{2\pi} \right)^2
 \left[  8  \log^2  \left( \frac{ \tau_0 }{\mu}   \right)    - 2  
  \log^2\left( \frac{ \tau_0 m e^{Y} }{\mu^2 } \right)
      - 2  \log^2 \left( \frac{\tau_0 m e^{-Y} }{\mu^2}  \right) 
    \right] \nn \\
 &&   \hspace{-3.ex} \times 
 \bigg[  \frac{e^Y}{m} \log\left( \frac{\tau_0 }{me^Y} \right) \,
\left[
2 \, x_1  \, \partial_{x_1} \,  
\right]   {\cal L}_{gg}      
+ \{x_1 \leftrightarrow x_2,  Y \leftrightarrow -Y \} \bigg]\,,   
\eea
for the $gg$ channel at the LL$_P$ accuracy 
and 
\bea 
\mathrm{d} {\hat \sigma}_{gq+qg\to H}^{(2)} &=&\mathrm{d} {\hat \sigma}_0\,
 \left( \frac{\alpha_s}{2\pi} \right)^2 \,
\Big( C_F  + C_A  \Big)  \, C_F  \bigg(
 \frac{e^Y}{m} {\cal L}_{gq}  \nn \\ 
  &\times & \log^2 \left(  \frac{ m e^Y }{\tau_0 }       \right) 
\log \left(
\frac{    \tau_0 e^Y    }{ m } 
\right)  +  \left\{ Y \leftrightarrow - Y\,, x_1 \leftrightarrow x_2  \right\}  \bigg) \,.
\eea
for the $gq + qg$ channel. Here we have normalized to the $gg$ channel color and spin average.

\section{Numerical results}\label{sect:numerics}

We now study the numerical consequences of the derived power corrections in the $N$-jettiness subtraction formalism. We focus on the Drell-Yan and gluon-fusion Higgs production channels at the LHC. The NNLO corrections to the Drell-Yan and $ggH$ processes have been implemented in {\tt MCFM v8.0} using $N$-jettiness subtraction~\cite{Boughezal:2016wmq}  and a thorough study of the impact due to the missing power corrections by comparing with the known exact results~\cite{Hamberg:1990np, Harlander:2003ai} has also been presented therein. In this section we follow exactly the same settings as used in Ref.~\cite{Boughezal:2016wmq} for $H/Z/W$ production at a 13 TeV LHC with NNLO the MSTW2008 PDF set~\cite{Martin:2009iq}. To study the impact of the power corrections calculated in the previous section, we generate the ${\cal O }(\alpha_s^2)$ NNLO coefficient using the $N$-jettiness subtraction scheme with and without the power corrections.

 \begin{figure*}
   \centering
  \includegraphics[width=0.9\linewidth]{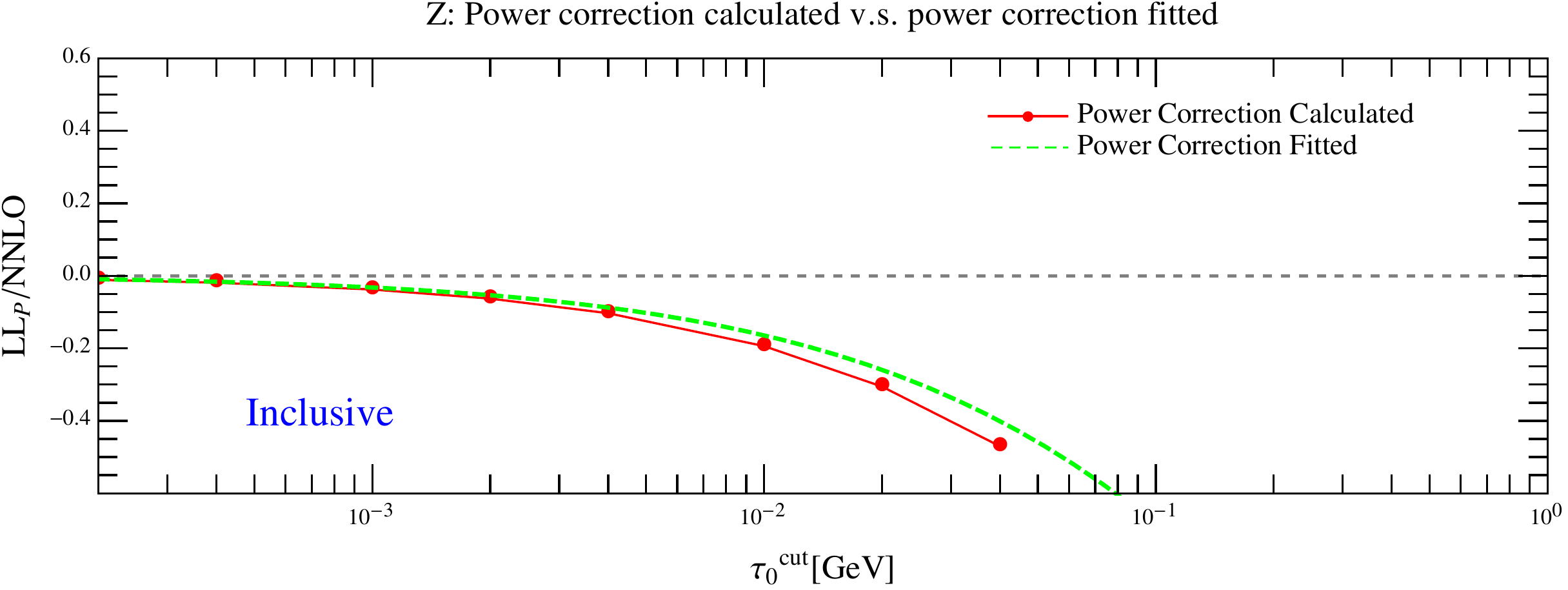}
  \includegraphics[width=0.9\linewidth]{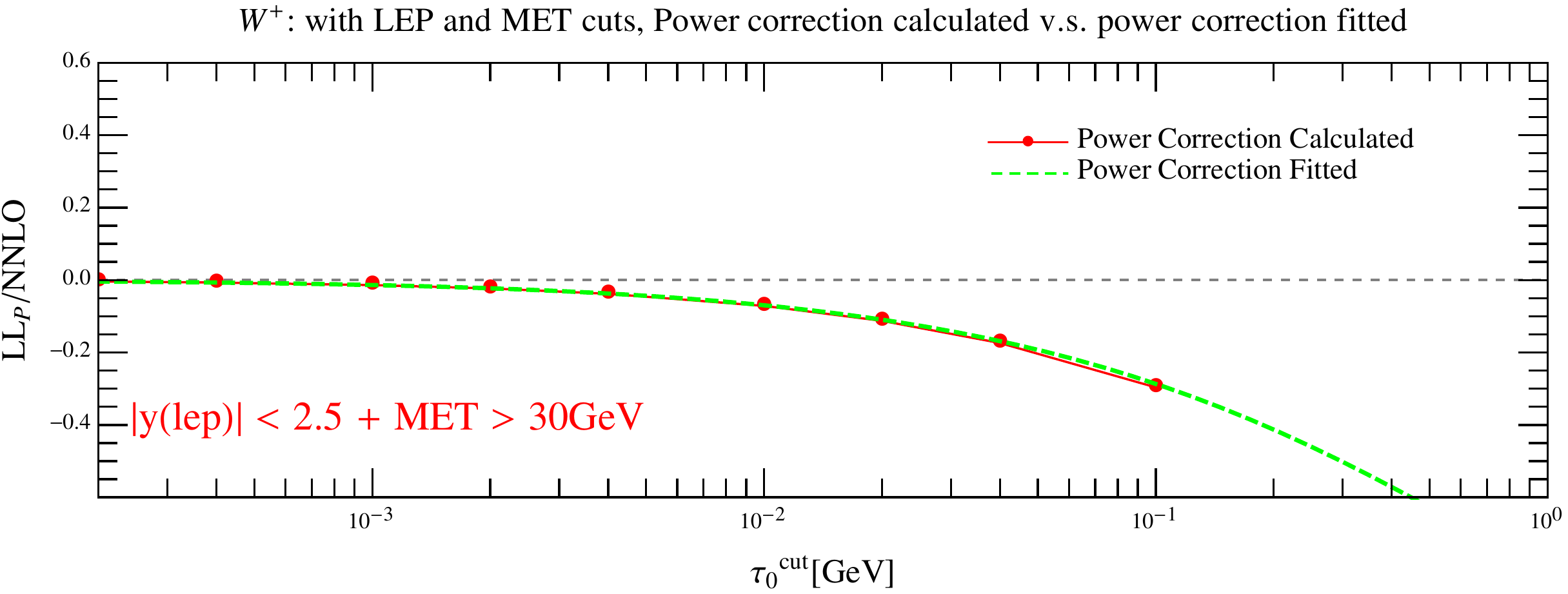}
 \caption{ A comparison of the fitted power corrections with the LL$_P$ calculated in this paper for inclusive $Z$-boson production (upper panel) and $W^+$-boson production with the indicated cuts on the final-state leptons (lower panel). We have normalized the ${\cal O}(\alpha_s^2)$ power corrections to the known ${\cal O}(\alpha_s^2)$ correction. }    
 \label{fig:check}
 \end{figure*}

We begin with a numerical validation of the calculated power corrections using $W$-boson and $Z$-boson production at a 13 TeV LHC with scale choice $\mu_R = 2m_V$ and $\mu_F = m_V/2$.  This scale choice is made to increase the size of the NNLO coefficient.  Due to the simple structure of the Drell-Yan cross section, the power corrections can be fitted to high accuracy by generating NNLO results with different $\tau_N^{cut}$ values, as has been first performed in Ref.~\cite{Boughezal:2016wmq}. In Fig.~\ref{fig:check}, we compare the calculated LL$_P$ (red solid line with dots)  at ${\cal O}(\alpha_s^2) $ with the fitted results (green dashed line) for both $Z$-boson and $W^+$-boson production.  We consider inclusive $Z$-boson production, and impose the following final-state cuts in the case of $W^+$ production:
\bea
y(lep) < 2.5,\;\;\; \not\!\! E_T > 30 \, \mathrm{GeV}.
\eea
From Fig.~\ref{fig:check}, we see that the calculated LL$_P$ and the fitted power corrections agree very well, and converge to each other in the small-$\tau_0$ region. The discrepancy seen in the larger $\tau_0$ region is due to the sub-leading logarithms which are included in the fit but not in the LL$_P$.

 \begin{figure*}
  \centering
  \includegraphics[width=0.9\linewidth]{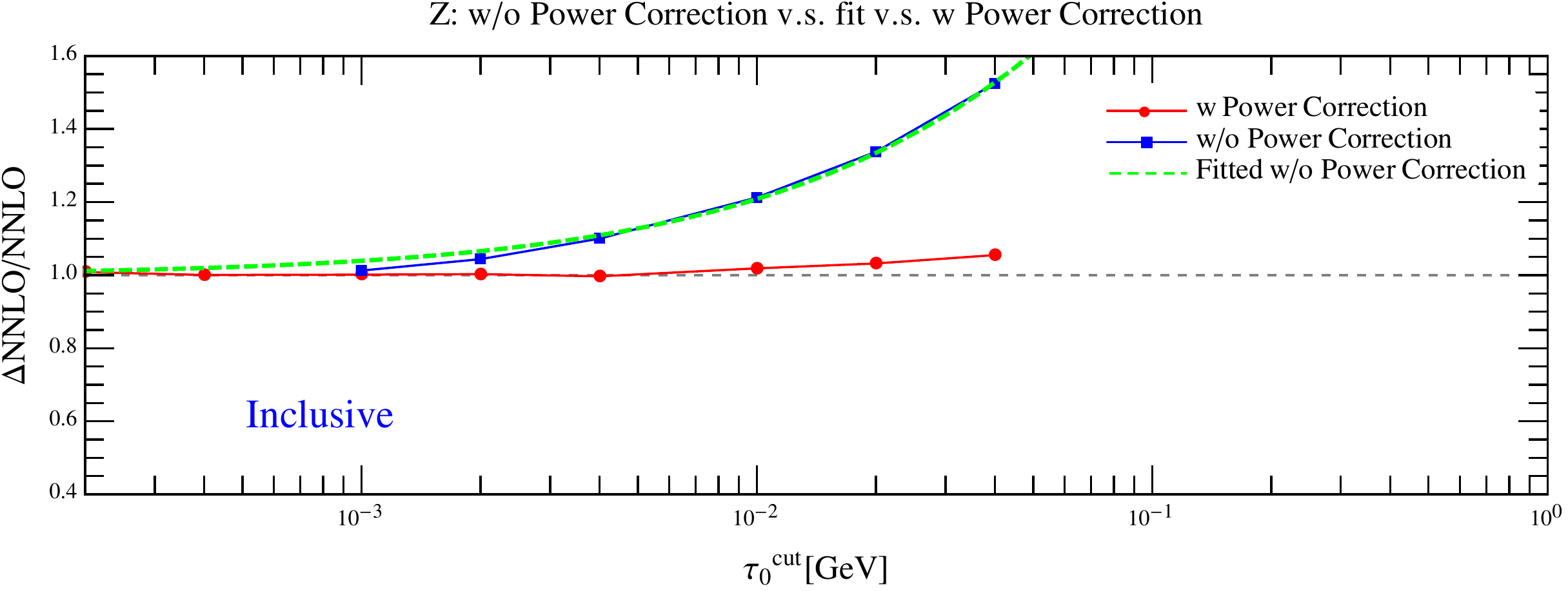}
  \includegraphics[width=0.9\linewidth]{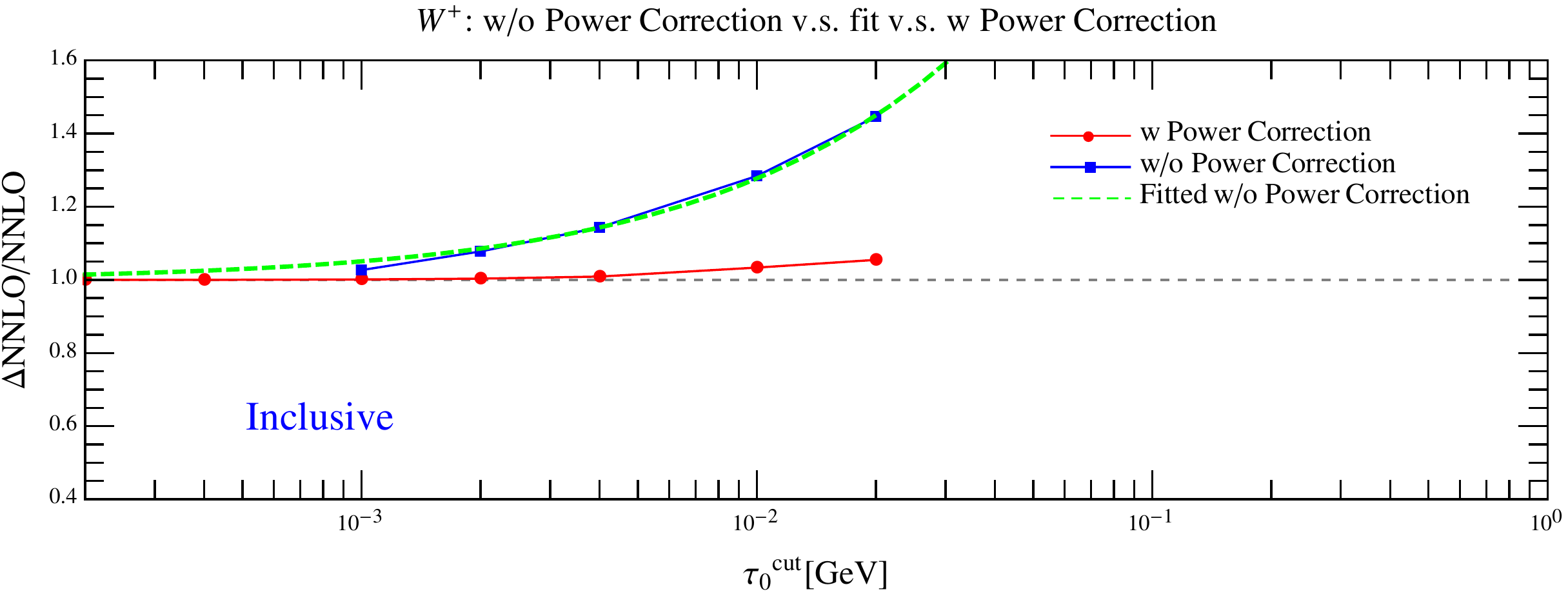}
 \caption{  The difference between the NNLO coefficients for inclusive $Z$-boson and $W^+$-boson production at the LHC using $N$-jettiness subtraction with and without power corrections, normalized to the known NNLO coefficient. We have plotted the difference between the $N$-jettiness result for the ${\cal O}(\alpha_s^2)$ correction and the known result, and have normalized this difference to the known correction, for this and all other figures. }    
 \label{fig:inclusive}
 \end{figure*}

Now we turn to the predictions for the NNLO coefficients for inclusive $Z/W$ production in which the impact of the power corrections are found to be relatively large~\cite{Boughezal:2016wmq}.   
 In Fig.~\ref{fig:inclusive}, we compare the $N$-jettiness subtraction scheme with and without the power corrections for both $Z$-boson production in the upper panel and $W^+$-boson production in the lower panel. 
We have plotted the difference between the ${\cal O}(\alpha_s^2)$ coefficient computed using $N$-jettiness and the exact ${\cal O}(\alpha_s^2)$ coefficient~\cite{Hamberg:1990np} and have normalized this to the known result. The vertical axis in fig.~\ref{fig:inclusive} characterizes the
deviation from the exact NNLO correction\footnote{We note that in Drell-Yan, a $15\%$ deviation in the ${\cal O}(\alpha_s^2)$ coefficient translates into a less than $1\%$ deviation in the total cross section }.
The blue solid line with squared dots represents the version without the power corrections generated using the current  {\tt MCFM v8.0}~\cite{Boughezal:2016wmq}.  The red line with round dots shows the results when adding on the analytically-calculated power corrections, while the green dashed line is the result obtained using the exact ${\cal O}(\alpha_s^2)$ coefficient subtracting out the power corrections fitted numerically. 
From Fig.~\ref{fig:inclusive}, we can see dramatic improvements in the convergence of  $N$-jettiness subtraction when the LL$_P$ is added. 
Without the power corrections,  $\tau_0^{cut}$ should be set to below $10^{-3} \, {\rm GeV}$ to reproduce the exact NNLO coefficients. The cut can be relaxed by a factor of $10$ when the power corrections are included.  As a consequence of the larger allowed $\tau_0$ cut, the computing time and the numerical efficiency are greatly improved. 

 \begin{figure*}
 \centering
  \includegraphics[width=0.9\linewidth]{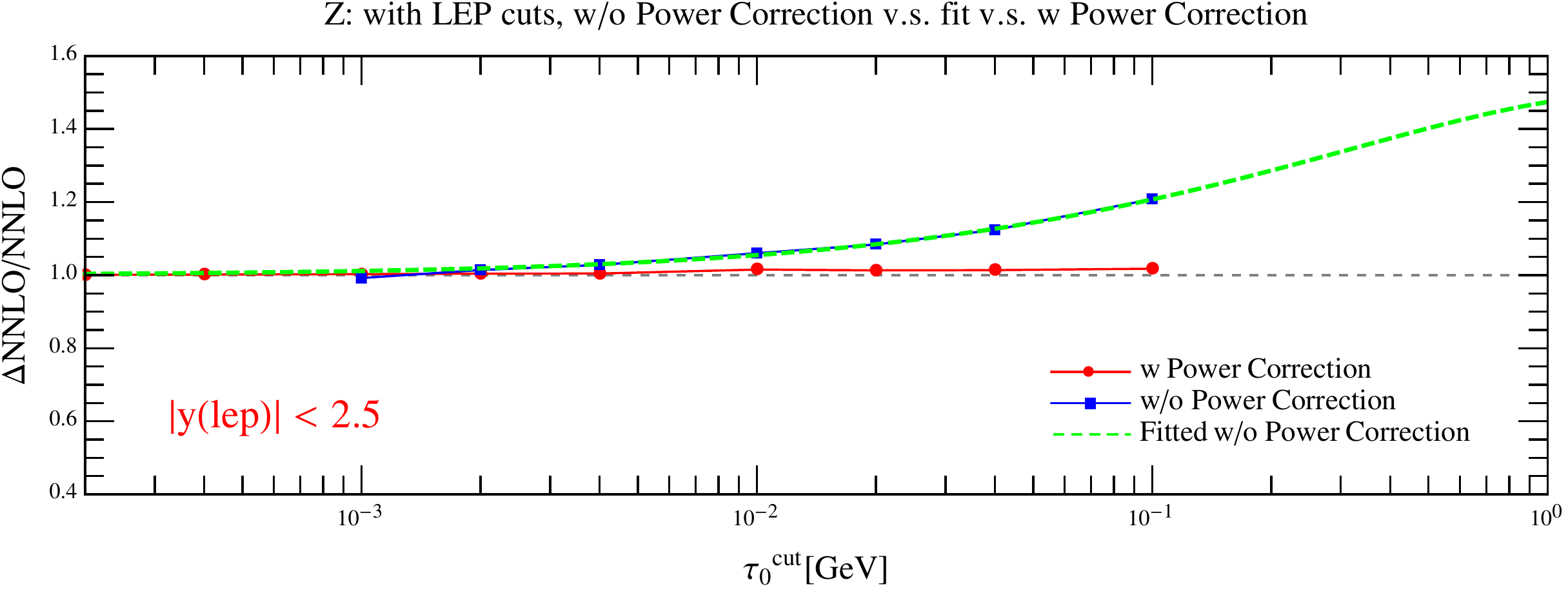} \vfill
  \includegraphics[width=0.9\linewidth]{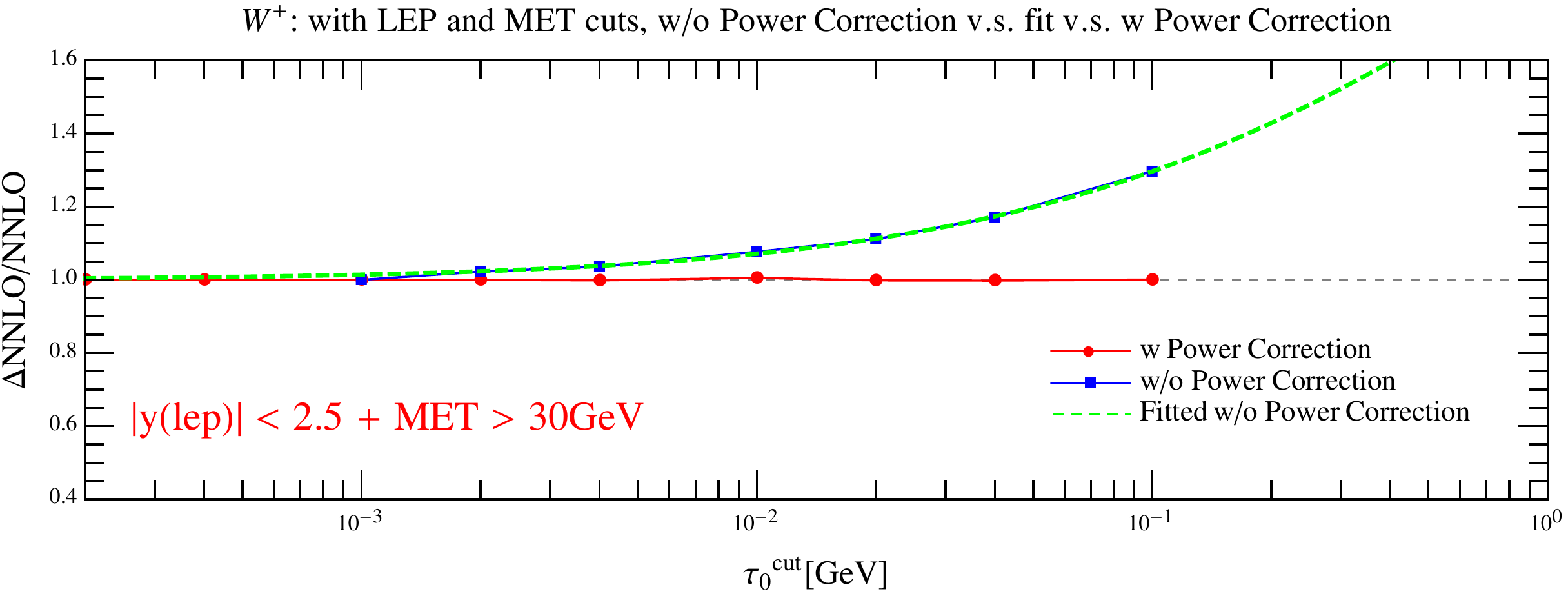}
 \caption{  The difference between the NNLO coefficients for $Z$-boson and $W^+$-boson production with lepton and missing energy cuts at the LHC obtained using $N$-jettiness subtraction
 with and without power corrections, normalized to the exact NNLO coefficients.  }    
 \label{fig:cuts}
 \end{figure*}

In Fig.~\ref{fig:cuts}, we show the comparison between $N$-jettiness subtraction with and without power corrections for $Z$-boson and $W^+$-boson production with cuts on the lepton rapidity, and on the missing energy in the case of the $W$-boson. With the presence of the cuts, the convergence is already better than in the inclusive case.  When the LL$_P$ power corrections are included, we see again a substantial improvement in the convergence of the $N$-jettiness subtraction scheme. At least a factor of $10$ in increasing $\tau_0^{cut}$ is observed, which leads to more efficient  realization of the NNLO calculation. 

\begin{figure*}
  \centering
  \includegraphics[width=0.9\textwidth]{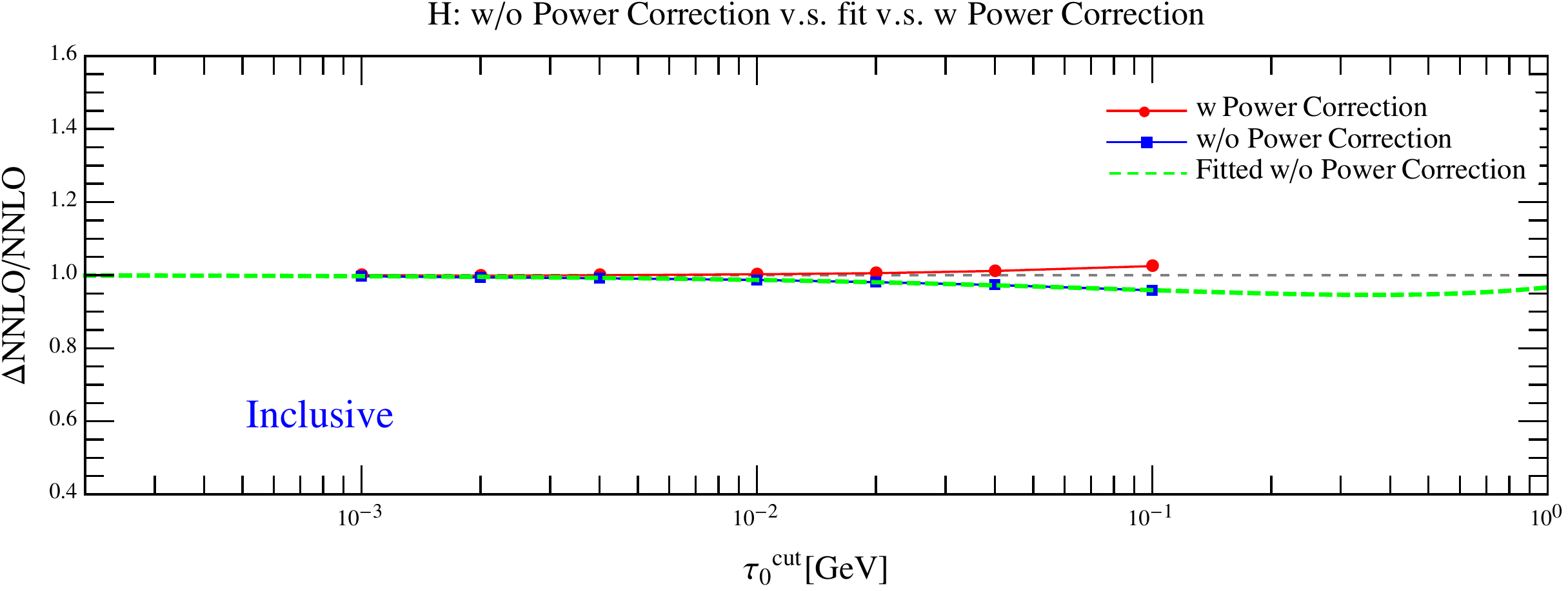}
 \caption{ A comparison of the NNLO coefficients for inclusive $ggH$ production at the LHC using $N$-jettiness subtraction
 with and without power corrections, normalized to the exact NNLO coefficient.   }    
 \label{fig:h-inclusive}
 \end{figure*}

Finally, in Fig.~\ref{fig:h-inclusive}, we consider gluon-fusion Higgs production at the LHC with scale choice $\mu_R = \mu_F = m_H$. Again we show the ${\cal O}(\alpha_s^2)$ coefficients both with power corrections included (blue solid line with squared dots) and without them(red solid line with round dots). We have normalized the results to the known exact NNLO coefficient~\cite{Harlander:2003ai}. In the $ggH$ case, the coefficients predicted without the power corrections already converge faster than the Drell-Yan case. Once the power corrections are included, similar improvement as in the Drell-Yan case  is also observed 
for gluon-fusion Higgs production.

\section{Summary}

In this manuscript we have studied the leading-logarithmic power corrections in the $N$-jettiness subtraction scheme. We have derived in detail the NLO power corrections for an arbitrary $N$-jet process, and have presented the important features of the derivation of the leading-logarithmic power corrections at NNLO for color-singlet production from both $q\bar{q}$ and $gg$ initiated processes.  The final expressions for the NNLO power corrections can be found in Section~\ref{sect:powernnlo}. We have found that for color-singlet production in hadronic collisions, the LL$_P$ at NNLO comes completely from the strongly-ordered soft limits. To get the LL$_P$ at NNLO, we only need the information from the NLO LL$_P$,  the leading-power collinear splitting kernels and the LO matrix elements. More interestingly the Abelian piece of the NNLO LL$_P$ is given by the convolution of the NLO leading logarithms in the leading power and the NLO LL$_P$. We note that a similar convolution structure also exists in the threshold case when the Drell-Yan threshold results are carefully studied~\cite{Hamberg:1990np}.
We conjecture that these features hold for the LL$_P$ for the production of an arbitrary number of jets. 

The final results for the LL$_P$, which are of the form $\alpha_s \log(\tau)$ at NLO and $\alpha_s^2 \log^3(\tau)$ at NNLO, are compact and can be easily added to existing numerical implementations of the $N$-jettiness subtraction scheme. Once the LL$_P$ are included, the resolution parameter $\tau_N^{cut}$ can be relaxed,  substantially improving the numerical convergence of the approach. We have demonstrated these improvements by studying both vector boson and Higgs production at the LHC in a variety of settings.  In all cases the incorporation of the LL$_P$ allows the value of $\tau_N^{cut}$ to be increased by nearly an order of magnitude while maintaining the same level of agreement with known results for color-singlet production at NNLO.

In the future, it will be important to complete the derivation of the NNLO power corrections for an arbitrary number of jets following the approach outlined here, as well as obtain the power corrections beyond LL$_P$  to further improve the efficiency of the $N$-jettiness subtraction scheme.  Other interesting directions included predicting and resumming power corrections within the framework of SCET~\cite{Kolodrubetz:2016uim,Pirjol:2002km}.

\section{Acknowledgments}
 
We thank A.~Isgr\`{o} for helpful conversations.  R.~B. is supported by the DOE contract DE-AC02-06CH11357.  F.~P. is supported by the DOE grants DE-FG02-91ER40684 and DE-AC02-06CH11357.  This research used resources of the Argonne Leadership Computing Facility, which is a DOE Office of Science User Facility supported under Contract DE-AC02-06CH11357. This research was also supported in part
by the NSF under Grant No. NSF PHY11-25915 to the
Kavli Institute of Theoretical Physics in Santa Barbara,
which we thank for hospitality during the completion of
this manuscript.

\smallskip
\noindent 
{\it Note added:} While we were finalizing this manuscript, Ref.~\cite{Moult:2016fqy} appeared, which calculates the NNLO leading-logarthmic power correction for color-singlet production through the $q\bar{q}$ partonic process.  Although their approach is different from ours, their results for Drell-Yan production fully agree with the LL$_P$ expressions presented here.
 
 \appendix
  
\section{The soft current for the $qg$ channel}

We list here the soft current for the $qg$ final state in Drell-Yan used to derive the LL$_P$ at NNLO. 
The $q_1(k_1) g_2(k_2) $ soft current can be derived by studying the leading-power NNLO splitting function~\cite{Catani:1999ss} and 
is given by the sum of the following five terms: 
\bea\label{eq:A1}
S^{(2)}_1 = \left[
(4\pi\alpha_s)  \, \mu^{2\epsilon} \, 2C_F \, \frac{{\hat s}}{ {\hat t}_2  \, {\hat u}_2 } \, 
\right] \,
\left[
(8\pi\alpha_s) C_F \, \mu^{2\epsilon} \,  \left(
\frac{-1 }{ {\hat u}_1     }
-
\frac{-1 }{   {\hat u}_1 +{\hat u}_2    }
\right)
\right] \,, 
\eea

\bea\label{eq:A2}
S^{(2)}_2 = 2(4\pi\alpha_s)^2 C_F^2 \left(
\frac{1}{ -\hat{u}_1  }
\,
\frac{{\hat t}_1}{  {\hat t}_2\,  \, 2 k_1 \cdot k_2}
\right) \,, 
\eea

\bea\label{eq:A3}
S^{(2)}_3 = 2(4\pi\alpha_s)^2 C_F^2 \mu^{4\epsilon}\left[
\frac{3}{-\hat{u}_1}  \, \frac{\hat{t}_1}{ \hat{t}_2  2 k_1 \cdot k_2 }
\,
+ 2 \frac{ - u_2 }{\hat{u}_1 + \hat{u}_2 }\frac{1}{-\hat{u}_1}  \frac{\hat{t}_1}{ \hat{t}_2  2 k_1 \cdot k_2 }
\,
- 2 \frac{- \hat{u}_2 }{ (\hat{u}_1 + \hat{u}_2)^2} \, \frac{\hat{u}_1}{\hat{u}_2\, 2 k_1 \cdot k_2 }
\,
\right] \,,  \quad\quad
\eea 

\bea\label{eq:A4}
S^{(2)}_4 = \left[ 
 2(4\pi\alpha_s)\, \mu^{2\epsilon}   C_A
 \frac{\hat{s} }{ \hat{t}_2 \,\hat{u}_2  } 
 \right] \,
\left[
(8\pi\alpha_s)\, \mu^{2\epsilon} 
C_F \, \left(
\frac{-1}{\hat{u}_1}
+ \frac{- 1}{2}\, \frac{-1}{\hat{u}_1}
- \frac{-1}{2} \, \frac{-1}{\hat{u}_1+\hat{u}_2}
\right)
\right] \,, \quad\quad
\eea

\bea\label{eq:A5}
S^{(2)}_5 = 2(4\pi\alpha_s)^2 \mu^{4\epsilon}  C_F \, C_A
\left[
\frac{\hat{u}_1}{  \hat{u}_2 \, 2  k_1 \cdot k_2     } \left(
- \frac{1}{\hat{u}_1} - \frac{   1  }{ \hat{  u }_1+ \hat{  u }_2}
\right)
- \, \left( \frac{1}{-\hat{u}_1}  
+ \frac{1}{-\hat{u}_1 - \hat{u}_2} \,
\right) \frac{{\hat t}_1}{ {\hat t}_2 \, 2k_1 \cdot k_2 }
\right] \,.  \quad\quad
\eea
Here $\hat{s} = m^2$, $\hat{t}_i = p_a \cdot k_i$ and $\hat{u}_i = p_b \cdot k_i $. 

The leading singular contribution in the strongly-ordered limit $E_g \ll E_q$ is given by
\bea\label{eq:A6}
S^{(2)}_{s.o.}  &=& S_{a1}^{(2)} + S_{qb}^{(2)} + S_{b1}^{(2)} \, \nn \\
&=&  
(4\pi\alpha_s) \, \mu^{2\epsilon} \, 2 \, \left[ 
 ( 2 \, C_F -  C_A) 
\,
\frac{{\hat t}_1}{  {\hat t}_2\,  \, 2 k_1 \cdot k_2} \, 
+
C_A \frac{\hat{s}}{\hat{t}_2 \, \hat{u}_2 }
+
C_A \frac{\hat{u}_1}{\hat{u}_2 \, 2 k_1 \cdot k_2}
\right] \,
\left[
(8\pi\alpha_s)\, \mu^{2\epsilon} 
C_F \,
\frac{1}{- \hat{u}_1}\right] \,, \quad\quad
\eea
with
\bea
&& S^{(2)}_{a1} = (4\pi\alpha_s) \, \mu^{2\epsilon} \, 2 \, \left[ 
 ( 2 \, C_F -  C_A) 
\,
\frac{{\hat t}_1}{  {\hat t}_2\,  \, 2 k_1 \cdot k_2} \, 
\right] \,
\left[
(8\pi\alpha_s)\, \mu^{2\epsilon} 
C_F \,
\frac{1}{- \hat{u}_1}\right] \,,  \nn \\
&& S^{(2)}_{ab} = (4\pi\alpha_s) \, \mu^{2\epsilon} \, 2 \, \left[ 
C_A \frac{\hat{s}}{\hat{t}_2 \, \hat{u}_2 }
\right] \,
\left[
(8\pi\alpha_s)\, \mu^{2\epsilon} 
C_F \,
\frac{1}{- \hat{u}_1}\right] \,,  \nn \\
&& S^{(2)}_{b1} = (4\pi\alpha_s) \, \mu^{2\epsilon} \, 2 \, \left[ 
  C_A \frac{\hat{u}_1}{\hat{u}_2 \, 2 k_1 \cdot k_2}
\right] \,
\left[
(8\pi\alpha_s)\, \mu^{2\epsilon} 
C_F \,
\frac{1}{- \hat{u}_1}\right] \,. 
\eea
We note that $S_{s.o.}^{(2)}$ factorizes into the product of a leading-power soft gluon current that knows about three eikonal directions, and a sub-leading soft quark current. 

The final results obtained after integrating Eq.~(\ref{eq:A1}) through Eq.~(\ref{eq:A5}) over $k_1$ and $k_2$  can be identified with the terms in Eq.~(\ref{eq:A6}) in a one-to-one manner. 
Performing the phase space integration over $S_1^{(2)}$, $S_2^{(2)}$ to $S_5^{(2)}$, gathering all the pieces and 
 splitting into different color factors according to Eq.~(\ref{eq:A6}), we find in the $qg$ final state for Drell-Yan, the LL$_P$ receives
contributions from a $2 C_F-C_A $  term:
\bea
I^{(2)}_{a1} =  - \frac{1}{ \epsilon ^3} \,
\left( 
\frac{\alpha_s }{2\pi}
\right)^2 \frac{e^Y}{m} \,  
\,
\left( 2C_F -  C_A \right)  T_R
\, 
\left[
 \left(\frac{m \tau_0 e^Y}{\mu ^2}\right)^{- \epsilon }  
-  \left(\frac{\tau_0 }{\mu }\right)^{-2 \epsilon } 
\right]  \,  \left(\frac{m \tau_0 e^Y}{\mu ^2}\right)^{- \epsilon } + \dots  \,,  \quad\quad
\,
\eea
Contributions from $C_A$-correlated terms (from matrix element contributions proportional to $(k_1 \cdot k_2)^{-1}$) are given by
\bea
I^{(2)}_{b1} &=& - \frac{1}{2\epsilon^3} \,
\left( 
\frac{\alpha_s }{2\pi}
\right)^2 \frac{e^Y}{m} \,  
\, 
C_A T_R 
\left[ 
\left(\frac{m \tau_0 e^Y}{\mu ^2}\right)^{- \epsilon }
- 
\left(\frac{\tau_0 }{\mu }\right)^{-2 \epsilon } 
\right]  \left(\frac{m \tau_0  e^Y}{\mu ^2}\right)^{- \epsilon } \nn \\
&+& 
\frac{1}{\epsilon^3 }
 \left( 
\frac{\alpha_s }{2\pi}
\right)^2 \frac{e^Y}{m} \,  
\, 
C_A T_R \, 
 \left[ \left(\frac{\tau_0 }{\mu }\right)^{-2 \epsilon } \left(\frac{m \tau_0 e^{-Y}}{\mu ^2}\right)^{-\epsilon }
 +\frac{1}{2} \left(\frac{\tau_0 }{\mu }\right)^{-2 \epsilon } \left(\frac{m \tau_0 e^Y}{\mu ^2}\right)^{-\epsilon }
 \right. \nn \\
&& \hspace{25.ex} \left.  -\left(\frac{m \tau_0 }{\mu ^2}\right)^{-2 \epsilon }-\frac{1}{2} \left(\frac{\tau_0 }{\mu }\right)^{-4 \epsilon }
 \right]  + \dots \,,  
\eea
while the contributions from $C_A$ non-correlated terms is
\bea\label{eq:Iij}
I^{(2)}_{ab}&= & - \,
 \frac{1}{2\epsilon^3} \, 
 \left( 
\frac{\alpha_s }{2\pi}
\right)^2 \frac{e^Y}{m} \,  
\,
 C_A T_R   \,
 \left[ 2 \left(\frac{m \tau_0 e^Y}{\mu ^2}\right)^{-\epsilon }-\left(\frac{\tau_0 }{\mu }\right)^{-2 \epsilon }\right] 
 \left[ \left(\frac{m \tau_0 e^Y}{\mu ^2}\right)^{-\epsilon }-\left(\frac{\tau_0 }{\mu }\right)^{-2 \epsilon }\right]  \nn \\
 &+&
  \frac{1}{2\epsilon^3} \,
  \left( 
\frac{\alpha_s }{2\pi}
\right)^2 \frac{e^Y}{m} \,  
 C_A T_R
\left[ 2 \left(\frac{\tau_0 }{\mu }\right)^{-2 \epsilon } \left(\frac{m \tau_0 e^{-Y}}{\mu ^2}\right)^{-\epsilon }
+\left(\frac{\tau_0 }{\mu }\right)^{-2 \epsilon } \left(\frac{m \tau_0 e^Y}{\mu ^2}\right)^{-\epsilon } \right. \nn \\
&& \hspace{25.ex}
\left. -2 \left(\frac{m \tau_0 }{\mu ^2}\right)^{-2 \epsilon }
-\left(\frac{\tau_0 }{\mu }\right)^{-4 \epsilon }\right] + \dots \,.   
\eea
The ellipsis in these equations denote terms that do not contribute to the final LL$_P$ and that have therefore been omitted. We have normalized the results
to the $q{\bar q} \to V$ color and spin average by multiplying with $\frac{N_C}{N_C^2-1}$, which turns an overall factor of $C_F$ to $T_R$.  We have obtained $I^{(2)}_{a1}$, $I^{(2)}_{b1}$ and $I^{(2)}_{ab}$ using Eq.~(\ref{eq:A1}) to Eq.~(\ref{eq:A5}). These calculations show that the net results of $I_{a1}^{(2)}$, $I_{b1}^{(2)}$ and $I_{ab}^{(2)}$   
can be derived instead using the strongly-ordered limit in Eq.~(\ref{eq:A6}) with $S_{a1}^{(2)}$, $S_{b1}^{(2)}$ and $S_{ab}^{(2)}$, respectively.

\section{Leading-logarithmic terms at  leading power}

The leading-logarithmic coefficients at leading power for the Drell-Yan process at the NLO are given by 
\bea\label{eq:LLLP}
I^{(1)}_{LL} = \frac{\alpha_s C_F}{2\pi}
 \left( -\frac{8}{\mu} \left[  \frac{\log\left( \tau_0/\mu \right)}{\tau_0/\mu} \right]_+ 
  + \frac{2 m e^Y}{\mu^2} \left[
  \frac{\log\left( \tau_0m e^{Y} /\mu^2 \right)}{\tau_0 m e^{Y}  /\mu^2}
  \right]_+
    + \frac{2 m e^{-Y}}{\mu^2} \left[
  \frac{\log\left( \tau_0m e^{-Y} /\mu^2 \right)}{\tau_0 m e^{-Y}  /\mu^2}
  \right]_+
  \right)
 \,, \quad\quad
\eea
where the dependence on the Born cross section has been suppressed. We note that the convolution gives
\bea
A \left[ \frac{\log(A\, x) }{A\, x} \right]_+ \otimes
\log(B \, x)
=  \frac{1}{2}\, \log^2(A\,x)  \, \log(B\,x) + \dots \,,
\eea
where we have only kept the leading-logarithmic contribution, as denoted by the ellipsis.

\end{document}